\documentclass[aip,jap,reprint]{revtex4-1}
\usepackage{graphicx}

\begin{document}


\title{Enhancement of the thermoelectric properties in doped FeSb$_2$ bulk crystals} 



\author{Kefeng Wang}
\affiliation{Condensed Matter Physics and Materials Science Department, Brookhaven National Laboratory, Upton, New York 11973 U.S.A.}
\author{Rongwei Hu}
\affiliation{Condensed Matter Physics and Materials Science Department, Brookhaven National Laboratory, Upton, New York 11973 U.S.A.}
\altaffiliation{Present address: Ames Laboratory US DOE and Department of Physics and Astronomy, Iowa State University, Ames, IA 50011, U.S.A.}
\author{John Warren}
\affiliation{Instrumentation Division, Brookhaven National Laboratory, Upton, New York 11973 U.S.A.}
\author{C. Petrovic}
\affiliation{Condensed Matter Physics and Materials Science Department, Brookhaven National Laboratory, Upton, New York 11973 U.S.A.}


\date{\today}

\begin{abstract}
Kondo insulator FeSb$_2$ with large Seebeck coefficient would have potential in thermoelectric applications in cryogenic temperature range if it had not been for large thermal conductivity $\kappa$. Here we studied the influence of different chemical substitutions at Fe and Sb site on thermal conductivity and thermoelectric effect in high quality single crystals. At $5\%$ of Te doping at Sb site thermal conductivity is suppressed from $\sim 250$ W/Km in undoped sample to about 8 W/Km. However, Cr and Co doping at Fe site suppresses thermal conductivity more slowly than Te doping, and even at 20$\%$ Cr/Co doping the thermal conductivity remains $\sim 30$ W/Km. The analysis of different contributions to phonon scattering indicates that the giant suppression of $\kappa$ with Te is due to the enhanced point defect scattering originating from the strain field fluctuations. In contrast, Te-doping has small influence on the correlation effects and then for small Te substitution the large magnitude of the Seebeck coefficient is still preserved, leading to the enhanced thermoelectric figure of merit ($ZT\sim 0.05$ at $\sim 100$ K) in Fe(Sb$_{0.9}$Te$_{0.1}$)$_2$.
\end{abstract}

\pacs{}

\maketitle 

\section{Introduction}

The revival of research on the solid state thermoelectric cooling and electrical power generation devices could be mainly attributed to their attractive features, such as long life, the absence of moving parts and emissions of toxic gases, low maintenance and high reliability.\cite{TE1,TE2} Present thermoelectric materials have relatively low energy conversion efficiency that can be evaluated by thermoelectric figure of merit $ZT=(S^2/\rho\kappa)\cdot T$, where $S$ is the Seebeck coefficient, $\rho$ is the electrical resistivity, $\kappa$ is the thermal conductivity and $T$ is the absolute temperature. Recent efforts to design materials with enhanced thermoelectric properties were mainly focused on reducing the lattice contribution to thermal conductivity by alloy scattering, superstructures or nanostructure engineering.\cite{phonon1, phonon2, phonon3} On the other hand, interest in the potential merits of electronic correlation effects was revived by the discovery of large Seebeck coefficients in transition metal compounds, such as FeSi and Na$_x$CoO$_2$.\cite{FeSi,NaCoO} In a Kondo insulator, localized \textit{f} or \textit{d} states hybridize with conduction electron states leading to the formation of a small hybridization gap with the large density of states (DOS) just below and above the gap. This resonance in DOS centered at about 2-3 $k_BT$ from the Fermi energy could induce very large Seebeck coefficient, as observed in FeSi. \cite{correlated1,correlated2,correlated3} Up to now, most of the effort has been concentrated on the high temperature range whereas current thermoelectric materials achieve poor thermoelectric efficiency at the cryogenic temperature range. Space science applications, cryocooling, and microelectronic superconducting quantum interference devices could be improved by a reliable effective solid state cooling.

Very recently, simply binary compound FeSb$_2$ with an orthorhombic structure has been characterized as an example of strongly correlated non-cubic Kondo insulator with 3d ions. \cite{kondo1,kondo2} Large value of Seebeck coefficient at 10 K and a record high thermoelectric power factor $S^2/\rho\sim 2300~\mu WK^{-2}cm^{-1}$ were observed. \cite{fesb,fesb1,fesb2,rongwei} This might imply the FeSb$_2$-based materials with narrow energy gaps and correlated bands could be good thermoelectrics at cryogenic temperatures. However, ZT is very low due to the very large thermal conductivity values.

It was shown that properties of Kondo insulators are sensitive to chemical substitution.\cite{fesbte, doping2, doping3,sun} Forming solid solution, i.e., maximizing the influence of point-defect scattering by creating large mass difference and elastic strain at the lattice sites, proved to be an efficient approach to suppress the thermal conductivity and enhance ZT of some thermoelectric materials, such as filled skutterudites materials Co(Sb$_{1-x}$As$_x$)$_3$, Co$_{1-x}$Ni$_x$As$_3$, and bismuth based materials Bi$_2$(Se/Sb/Te)$_3$. \cite{CoSbAs,CoNiAs,bismuth1,bismuth2,bismuth3} Here, we clarify the influence of the different chemical substitution at Fe and Sb site on the thermoelectric properties of FeSb$_2$. By $5\%$ Te doping on Sb sites thermal conductivity reduces from $\sim 250$ W/Km in FeSb$_2$ to about 8 W/Km in Fe(Sb$_{0.95}$Te$_{0.05}$)$_2$. This is more significant than the suppression of thermal conductivity by Co or Cr doping at Fe site due to the different bonding tendency of Sb and Te. Whereas the doping at Fe site rapidly smears out the Seebeck coefficient peak at low temperature, for Te substitution at Sb site this peak survives in the wide doping region. Consequently an enhanced thermoelectric figure of merit ($ZT\sim0.05$ at $\sim 100$ K) in Fe(Sb$_{0.9}$Te$_{0.1}$)$_2$ is obtained when compared to $ZT<0.005$ in undoped FeSb$_2$.

\section{Experimental}

Fe$_{1-x}$T$_x$Sb$_2$ (T=Cr and Co) and Fe(Sb$_{1-x}$Te$_x$)$_2$ single crystals were grown from excess Sb flux, as described previously. \cite{fesbte,doping2,doping3} Powder X-ray diffraction (XRD) spectra of the ground samples were taken with Cu $K\alpha$ radiation ($\lambda=1.5418 {\AA}$) using a Rigaku Miniflex x-ray machine. The lattice parameters were obtained by fitting the XRD pattern using the RIETICA software. \cite{rietica} Elemental analysis was performed using an energy-dispersive x-ray spectroscopy (EDX) in a JEOL JSM-6500 scanning electron microscopy.
Thermal transport properties including thermal conductivity, Seebeck coefficient, as well as electrical resistivity were measured in Quantum Design PPMS-9 from 2 K to 300 K using one-heater-two-thermometer method. The direction of heat and electric current transport was along the c-axis of single grain crystals oriented using a Laue camera.

\section{Results and discussion}

Powder X-ray diffraction patterns show that all Fe(Sb$_{1-x}$Te$_x$)$_2$ ($x\leq 0.1$), (Fe$_{1-x}$Cr$_x$)Sb$_2$ and (Fe$_{1-x}$Co$_x$)Sb$_2$ samples crystallize in the \textit{Pnnm} structure without any impurity phases. For example, lattice parameters of Te-doped samples are shown in Fig. 1(a).  The effect of Te substitution on the Sb site is to expand the unit cell volume when compared to FeSb$_2$. This expansion is anisotropic and the lattice contracts in the \textit{ab}-plane while expands along the \textit{c}-axis (Fig. 1(a)). The elemental distribution by EDX elemental analysis revealed that the samples are homogenous (Fig. 1(b-e)) and that there is very small difference between the nominal and actual doping contents, as shown in Table I.

\begin{figure}
\includegraphics[scale=0.45]{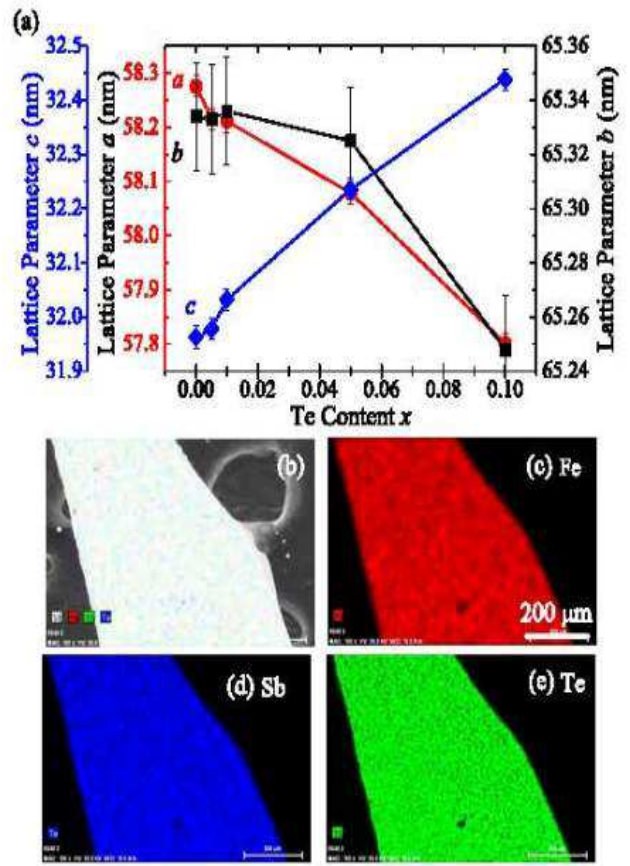}
\caption{(Color online) (a) Lattice constants  of samples versus nominal Te concentration $x$. (b-e) SEM image of a typical Te-doped crystal (b), and elemental distribution of Fe (c), Sb (d) and Te (e) by EDX elemental analysis in crystal with $x=0.1$. The scale of these figure are same as shown in (c).\label{}}
\end{figure}

Fig. 2 shows the temperature dependence of the thermal conductivity, Seebeck coefficient and electrical resistivity of crystals with different Te doping level at Sb site. Thermal conductivity $\kappa(T)$ achieves its maximum between 12 K and 20 K for pure FeSb$_2$, peaking at $\sim$ 250 W/Km. Seebeck coefficient changes sign from positive to negative around 120 K, indicating presence of two carrier types. The absolute value of $S$ increases rapidly below 40 K, and reaches its nearly constant maximum value $|S(T)|_{max}\sim800 \mu$V/K in the temperature interval $10$ K$ \sim ~20$ K.\cite{rongwei} Due to large thermal conductivity up to $\sim 200$ W/Km at 20 K, $ZT$ ($ZT<0.005$ at 10 K) is very low, limiting the prospect for applications. By $0.5 \%$ substitution of Te at Sb site, thermal conductivity is suppressed by half, i.e., $\kappa_{max}\sim90$ W/Km. Further doping with Te induces futher decrease in $\kappa$. At $5\%$ Te substitution thermal conductivity decreases to about 8 W/Km. The peak in thermal conductivity shifts to higher temperature with increase in Te content. Tellurium substitution also induces the suppression of resistivity and a change to metallic ground state (Fig. 2(c)).\cite{fesbte} The metallic temperature region increases with Te doping.

\begin{figure}
\includegraphics[scale=1.0]{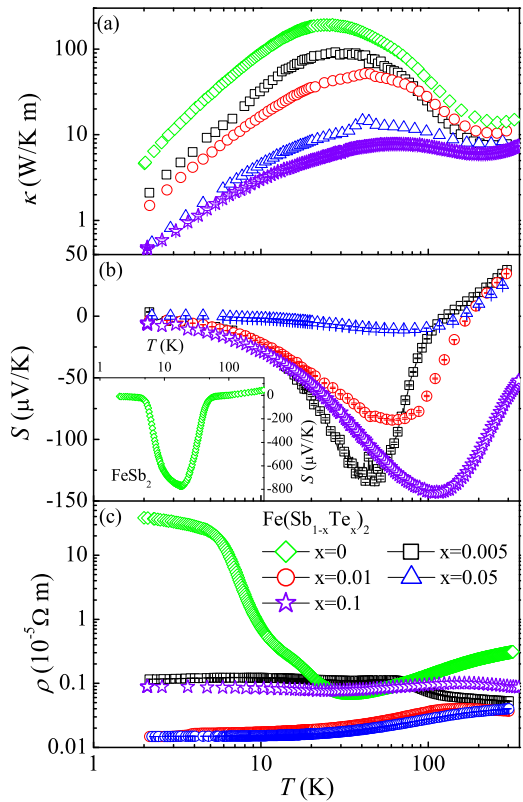}%
\caption{(Color online) Temperature dependence of thermal conductivity $\kappa$ (upper figure), Seebeck coefficient $S$ (middle figure) and resistivity $\rho$ (bottom figure) of Fe(Sb$_{1-x}$Te$_x$)$_2$ samples with $x=0, 0.001, 0.01, 0.05$ and 0.1, respectively.\label{}}%
\end{figure}

\begin{figure}
\includegraphics[scale=0.65]{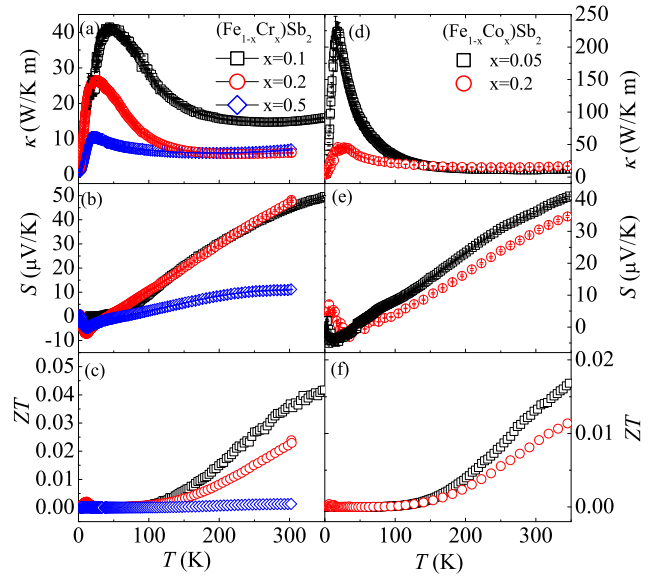}
\caption{(Color online) Temperature dependence of thermal conductivity $\kappa$ (a, d), Seebeck coefficient $S$ (b, e) and thermoelectric figure of merit $ZT$ (c, f) for (Fe$_{1-x}$Cr$_x$)Sb$_2$ (a-c) and (Fe$_{1-x}$Co$_x$)Sb$_2$ (d-f) crystals, respectively.}
\end{figure}

\begin{figure}
\includegraphics[scale=0.65]{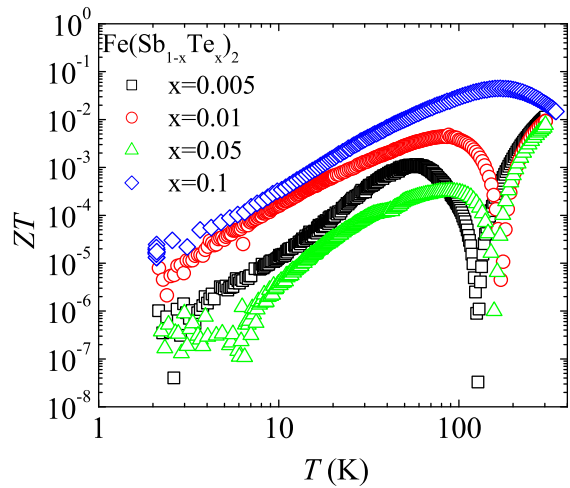}
\caption{(Color online) Thermoelectric figure of merit of all Te-doped crystals as a function of temperature.\label{}}%
\end{figure}

Chromium and cobalt substitution on Fe site reduces the thermal conductivity more slowly than Te doping at Sb site. About $5\%$ Co doping on Fe sites has nearly no influence on thermal conductivity, and $20\%$ Co and Cr doping reduces thermal conductivity to $\sim 30$ W/Km which is similar to the value observed in $1\%$ Te-doped sample (Fig. 3(a) and (d)). For $50\%$ Cr-doped sample the thermal conductivity decreases to $\sim 10$ W/Km comparable to the value in $5\%$ Te-doped sample.

The low-temperature peak of Seebeck coefficient is almost completely suppressed by about $5\%$ Co substitution remaining at $\sim -10$ $\mu$V/K for all samples with Cr/Co doping level from $5\%$ to $50\%$ (Fig. 3 (b) and (c)). Hence $ZT$ value are very small (Fig. 3(c) and (f)). In contrast, Seebeck coefficient is also suppressed but the peak value at low temperature remains large in Te-doped system. The peak value of Seebeck coefficient decreases from $\sim 800$ $\mu$V/K for $x=0$ to  $\sim 200$ $\mu$V/K for $x=0.005$, and shifts to high temperature. $ZT$ values for Te-doped crystals are shown in Fig. 4. The maximum of $ZT$ in Fe(Sb$_{0.9}$Te$_{0.1}$)$_2$ reaches $\sim 0.05$ at about 100 K, which is nearly one-order of magnitude larger than $ZT\sim 0.005$ in undoped FeSb$_2$.

The phonon drag mechanism is not likely to contribute significantly to $S$ since isostructural RuSb$_2$ and FeAs$_2$ have larger $\kappa$ and smaller peak values of $S$.\cite{fesb1,fesb2} It is pointed out that the large thermopower in FeSb$_2$ originates from correlated electron effects in bands with the high density of states near the gap edges.\cite{correlated1} Therefore, Co/Cr substitution on Fe sites have more significant influence on the electronic correlations.

\begin{figure}
\includegraphics[scale=1.0]{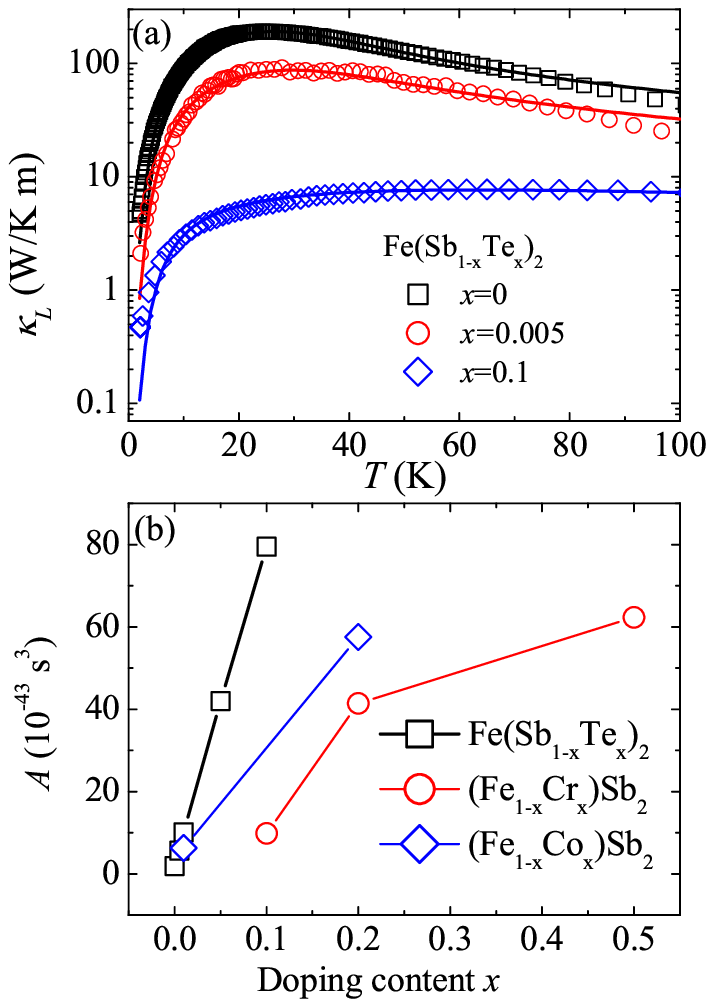}
\caption{(Color online) (a) Experimental (symbols) and calculated (lines) lattice thermal conductivity $\kappa_L$ for Fe(Sb$_{1-x}$Te$_x$)$_2$ samples with $x=0, 0.005$ and 0.1, respectively. (b) Point defect scattering rate coefficient $A$ vs doping concentration for crystals with Te, Cr and Co doping. \label{}}%
\end{figure}

\begin{table}[b]
\caption{Fitting parameters L, A and B for the lattice thermal conductivity of doped FeSb$_2$ samples using Eqs. (1) and (2).}
\begin{ruledtabular}
\begin{tabular}{ccccccc}
Nominal \textit{x} & Actual \textit{x} & L ($\mu$m) & A ($10^{-43}$ s$^3$)& B (10$^{-18}$ sK$^{-1}$)\\
Fe(Sb$_{1-x}$Te$_x$)$_2$ \\
0 & - & 72 & 1.9 & 2.4 \\
0.005 & 0.006(7) & 23 & 5.6 & 3.8\\
0.01 & 0.007(9) & 51 & 10.1 & 2.1 \\
0.05 & 0.04(3) & 42 & 40.7 & 3.5 \\
0.1 & 0.09(1) & 67 & 79.5 & 2.7 \\
(Fe$_{1-x}$Cr$_x$)Sb$_2$ \\
0.1 & 0.08(3) & 82 & 9.8 & 1.9 \\
0.2 & 0.17(5) & 21 & 41.4 & 5.8 \\
0.5 & 0.04(4) & 37 & 62.3 & 7.4 \\
 (Fe$_{1-x}$Co$_x$)Sb$_2$ \\
0.01 & 0.009(1) & 31 & 6.3 & 4.4 \\
0.2 & 0.18(7) & 22 & 57.5 & 2.3
\end{tabular}
\end{ruledtabular}
\end{table}

Thermal conductivity is composed of the electron term $\kappa_e$ and the phonon term $\kappa_{ph}$; $\kappa_{total}=\kappa_e+\kappa_{ph}$. The electron term $\kappa_e$ estimated using Wiedemann-Franz law $\frac{\kappa_e}{T}=\frac{L_0}{\rho}$, is $0.1\%$ of the total thermal conductivity, indicating a predominantly phonon contribution. The suppression of thermal conductivity should reflect enhanced phonon scattering. In general, phonon scattering can be realized through point defect scattering, carrier-phonon scattering, boundary scattering, phonon Umklapp scattering and void filling.\cite{thermalconductivity1,thermalconductivity2,thermalconductivity3} In what follows we do not consider the carrier-phonon scattering processes since the carrier concentrations in our crystrals are low. Our samples are single crystals, hence the boundary scattering should not dominate the phonon scattering process. Moreover, Umklapp process should not vary significantly by replacing small amount of Sb with Te. Thus only enhanced point defect scattering  should be responsible for the suppression of thermal conductivity.

The lattice thermal conductivity is usually treated using the Debye approximation:\cite{thermalconductivity2,thermalconductivity4}
\begin{eqnarray}
\kappa_L=\frac{k_B}{2\pi^2\upsilon}\left(\frac{k_B}{\hbar}\right)^3T^3\int^{\frac{\theta_D}{T}}_0 \frac{\tau_cx^4e^x}{(e^x-1)^2}dx,
\end{eqnarray}
where $x=\frac{\hbar\omega}{k_BT}$ is dimensionless, $\omega$ is the phonon frequency, $k_B$ is the Boltzmann constant, $\hbar$ is the Planck constant, $\theta_D$ is the Debye temperature, $\upsilon$ is the velocity of sound, and $\tau_c$ is the relaxation time.\cite{thermalconductivity2,thermalconductivity4} The overall relaxation rate $\tau_c^{-1}$ can be determined by combining various scattering processes
\begin{eqnarray}
\tau_c^{-1} &=&\tau_B^{-1}+\tau_D^{-1}+\tau_U^{-1} \\
&=&\frac{\upsilon}{L}+A\omega^4+B\omega^2Te^{-\frac{-\theta_D}{3T}},
\end{eqnarray}
where $\tau_B, \tau_D$ and $\tau_U$ are the relaxation times for boundary scattering, defect scattering and Umklapp processes, respectively, $A$ is the Rayleigh point defect rate.

The lattice thermal conductivity of all crystals can be well understood with this model. For example, Fig. 5(a) shows the lattice thermal conductivity (symbols) and fitting results using Eqns. 1 and 2 (lines) for Te-doped samples Fe(Sb$_{1-x}$Te$_x$)$_2$ with $x=0,0.005$ and 0.1. The fitting parameters for all samples are shown in Table I. The fitted grain size varies from 42 to 80 $\mu$m with no evident trend among the samples. The Umklapp scattering parameters $B$ are small and do not exhibit significant change with doping. The point defect scattering rate $A$ increases with increase in doping level, as shown in Fig. 5(b) and Table I. This confirms that the suppression of thermal conductivity in doped FeSb$_2$ originates from enhanced point defect scattering. The enhanced point defect scattering in solid solutions has been reported in many semiconductor systems, such as doped CoSb$_3$, Bi$_2$(Se/Sb/Te)$_3$, Si-Ge solution.\cite{CoSbAs,CoNiAs,bismuth1,bismuth2,bismuth3} The suppression of thermal conductivity by Te doping at Sb site is more significant when compared to these solutions and Co/Cr-doped FeSb$_2$. 

The prefactor $A$ for point defect scattering can be written as $A=\Omega_0\Gamma/(4\pi\nu^3)$ where $\Omega_0$ is the unit cell volume, $\nu$ is the sound velocity and $\Gamma$ is the scattering parameter.\cite{klemens,thermalconductivity1} In solid solutions, point defect scattering originates from the mass fluctuations ($\Gamma_M$) and interatomic coupling force difference (strain field fluctuations)($\Gamma_S$). These two contribution are additive: $\Gamma=\Gamma_M+\Gamma_S$. For the solution Fe(Sb$_{1-x}$Te$_x$)$_2$, $\Gamma_M=(2/3)(M_{Sb,Te}/M)^2x(1-x)(\Delta M/M_{Sb,Te})^2$.\cite{slack2} Since the mass difference between Sb and Te is very small (only $\sim 6 \%$) and is nearly same to the relative change between Fe and Co/Cr, the mass fluctuation can not induce large increase in point defect scattering ratio. Therefore, mechanism of point defect scattering most likely originates from the strain field fluctuation due to differences in size or bonding. The size difference between Sb and Te is also close to the difference btween Fe and Co/Cr. According to Pauling electronegativity \cite{pauling}, Fe, Co and Cr have nearly same electronegativity, whereas Sb and Te have different electronegativity and reside at different side of the metal/nometal boundary. This results in more than one order of larger $\kappa$ at room temperature and nearly 3 orders of magnitude at 10 K for Sb when compared to Te. \cite{SbTe1,SbTe2,SbTe3} Subsequently the Te doping and Fe-Te bonds likely induce large bonding fluctuation which is absent in Co/Cr-doped systems. This induces the large strain field fluctuation and enhanced point defect scattering in Te-doped FeSb$_2$.

\section{Conclusion}

In summary, we studied the influence of different chemical substitution at Fe and Sb site on thermal conductivity and thermoelectric effect of FeSb$_2$ in high quality single crystals. All dopants suppress the thermal conductivity. At $5\%$ of Te doping at Sb site thermal conductivity is suppressed from $\sim 250$ W/Km in undoped sample to about 8 W/Km. However, Cr and Co doping at Fe site suppress thermal conductivity more slowly than Te doping, and even at 20$\%$ Cr/Co doping the thermal conductivity remains $\sim 30$ W/Km. The analysis of different contributions to phonon scattering indicates that the giant suppression of $\kappa$ is due to the enhanced point defect scattering originating from the mass and strain field fluctuations. The significant enhancement of strain field fluctuation and suppression of thermal conductivity by Te doping is attributed to the different bonding tendency and thermal conductivity variations for Sb and Te. On the other hand, the doping at Fe site smears out the thermopower peak at low temperature significantly, while for Te doping at Sb site this peak survives in a wide doping region.  Consequently thermoelectric figure of merit $(ZT\sim0.05)$ in Fe(Sb$_{0.9}$Te$_{0.1}$)$_2$ at $\sim 100K$ is enhanced by one order of magnitude when compared to FeSb$_2$.

\begin{acknowledgments}
This work was carried out at Brookhaven National Laboratory. Work at Brookhaven is supported by the U.S. DOE under contract No. DE-AC02-98CH10886.
\end{acknowledgments}


%

%




\end{document}